\documentclass{elsart}

\usepackage{natbib}

\usepackage{amssymb}

\usepackage{graphicx}

\def\mbox{\hbox}           

\def\deg{\ifmmode ^\circ                
         \else $^\circ$
         \fi
         \hskip -0.1truecm}
\def\degd#1.#2{                         
               \ifmmode {#1^{\hskip 0.05em\circ}\hskip-0.42em.\hskip0.08em#2}
               \else {#1$^{\hskip 0.05em\circ}\hskip-0.42em.\hskip0.08em$#2}
               \fi
              }
\def\mind#1.#2{                         
               \ifmmode {#1^{\hskip 0.05em\prime}\hskip-0.35em.\hskip0.05em#2}
               \else {#1$^{\hskip 0.05em\prime}\hskip-0.35em.\hskip0.05em$#2}
               \fi
              }
\def\secd#1.#2{                         
               \ifmmode {#1^{\prime\prime}\hskip-0.46em.\hskip0.12em#2}
               \else {#1$^{\prime\prime}\hskip-0.46em.\hskip0.12em$#2}
               \fi
              }
\def\timsecd#1.#2{                      
                  \ifmmode {#1^{\rm s}\hskip-0.39em.\hskip0.08em#2}
                  \else {$#1^{\rm s}\hskip-0.39em.\hskip0.08em#2$}
                  \fi
                 }
\def\hms#1h#2m#3s{                      
                  \relax
                  \ifmmode #1^{\rm h}\,#2^{\rm m}\,#3^{\rm s}
                  \else \hbox{$#1^{\rm h}\,#2^{\rm m}\,#3^{\rm s}$}
                  \fi
                 }
\def\dms#1d#2m#3s{                      
                  \relax
                  \ifmmode #1^\circ\,#2^{\prime}\,#3^{\prime\prime}
                  \else \hbox{$#1^\circ\,#2^{\prime}\,#3^{\prime\prime}$}
                  \fi
                 }
\def\dmsd#1d#2m#3.#4s{                  
                      \relax
                      \ifmmode #1^\circ\,#2^{\prime}\,#3^{\prime\prime}
                               \hskip-0.46em.\hskip0.12em#4
                      \else \hbox{$#1^\circ\,#2^{\prime}\,#3^{\prime\prime}
                            \hskip-0.46em.\hskip0.12em#4$}
                      \fi
                     }
\def\hm#1h#2m{                          
              \relax
              \ifmmode #1^{rm h}\,#2^{\rm m}
              \else \hbox{$#1^{\rm h}\,#2^{\rm m}$}
              \fi
             }
\def\dm#1d#2m{                          
              \relax
              \ifmmode #1^\circ\,#2^{\prime}
              \else \hbox{$#1^\circ\,#2^{\prime}$}
              \fi
             }
\def\hmsd#1h#2m#3.#4s{                  
                      \relax
                      \ifmmode #1^{\rm h}\,#2^{\rm m}\,#3^{\rm s}
                               \hskip-0.39em.\hskip0.08em#4
                      \else \hbox{$#1^{\rm h}\,#2^{\rm m}\,#3^{\rm s}
                            \hskip-0.39em.\hskip0.08em#4$}
                      \fi
                     }
\def\hmd#1h#2.#3m{                  
                  \relax
                  \ifmmode #1^{\rm h}\,#2^{\rm m}
                           \hskip-0.55em.\hskip0.22em#3
                  \else \hbox{$#1^{\rm h}\,#2^{\rm m}
                        \hskip-0.55em.\hskip0.22em#3$}
                  \fi
                 }
\def\mg{\relax                          
        \ifmmode ^{\rm m}
        \else $^{\rm m}$
        \fi
       }
\def\mgd#1.#2{                          
              \relax
              \ifmmode #1^{\rm m}
                       \hskip-0.55em.\hskip0.22em#2
              \else \hbox{#1$^{\rm m}
                    \hskip-0.55em.\hskip0.22em$#2}
              \fi
             }

\def\la{\mathrel{\hbox{\rlap{\hbox{\lower4pt\hbox{$\sim$}}}\hbox{$<$}}}}
\def\ga{\mathrel{\hbox{\rlap{\hbox{\lower4pt\hbox{$\sim$}}}\hbox{$>$}}}}

\def\unitspace{\;}                      

\def\un#1{\ifmmode \unitspace\mbox{\rm #1} 
          \else $\unitspace$#1
          \fi}
\def\pun#1#2{\ifmmode \unitspace\mbox{\rm #1}^{#2} 
             \else $\unitspace$#1$^{#2}$
             \fi}

\def\kms{\un{km}\pun{s}{-1}}          
\def\Lsun{\ifmmode \un{L}_{\odot}     
          \else $\un{L}_{\odot}$
          \fi}
\def\Msun{\ifmmode \un{M}_{\odot}     
          \else $\un{M}_{\odot}$
          \fi}
\def\mum{\ifmmode \unitspace\mu\mbox{\rm m} 
         \else $\unitspace\mu$m
         \fi}
\def\sqarcsec{\ifmmode \unitspace\Box''    
              \else $\unitspace\Box''$     
              \fi} 

\def\Bp{\relax                            
        \ifmmode B_{||}                   
        \else $B_{||}$
        \fi}
\def\Bt{\relax                            
        \ifmmode B\!_{\perp}              
        \else $B\!_{\perp}$               
        \fi}
\def\Gcr{\relax                           
         \ifmmode \Gamma\!_{\rm cr}       
         \else $\Gamma\!_{\rm cr}$
         \fi}
\def\ICII{\relax                          
          \ifmmode I_{[\CII]}             
          \else $I_{[\CII]}$
          \fi}
\def\LHtwo{\relax                                 
           \ifmmode L_{\mbox{\rm\scriptsize H}_2} 
           \else $L_{\mbox{\rm\scriptsize H}_2}$  
           \fi}
\def\LLya{\relax                          
          \ifmmode L_{{\rm Ly}\,\alpha}   
          \else $L_{{\rm Ly}\,\alpha}$
          \fi}
\def\MHtwo{\relax                                 
           \ifmmode M_{\mbox{\rm\scriptsize H}_2} 
           \else $M_{\mbox{\rm\scriptsize H}_2}$  
           \fi}
\def\MHtwodot{\relax                                       
              \ifmmode \dot{M}_{\mbox{\rm\scriptsize H}_2} 
              \else $\dot{M}_{\mbox{\rm\scriptsize H}_2}$  
              \fi}                                         
\def\Mstardot{\relax                      
              \ifmmode \dot{M}_{\ast}     
              \else $\dot{M}_{\ast}$      
              \fi}
\def\nHI{\relax                                      
         \ifmmode n_{\mbox{\scriptsize\rm H\,\sc I}} 
         \else $n_{\mbox{\scriptsize\rm H\,\sc I}}$
         \fi}
\def\nHtwo{\relax                                
           \ifmmode n_{{\mbox{\scriptsize H}}_2} 
           \else $n_{{\mbox{\scriptsize H}}_2}$  
           \fi}
\def\rhostardot{\relax                         
                \ifmmode \dot{\rho}_{\ast}     
                \else $\dot{\rho}_{\ast}$      
                \fi}
\def\rhoZdot{\relax                          
             \ifmmode \dot{\rho}_{\rm Z}     
             \else $\dot{\rho}_{\rm Z}$      
             \fi}

\def\sou#1#2{\relax                       
             \ifmmode {\rm #1}\,{\rm #2}  
             \else #1$\,$#2
             \fi}

\def\NGC#1{\sou{NGC}{#1}}                

\def\qu#1#2{\relax                          
            \ifmmode #1_{\rm #2}            
            \else $#1_{\rm #2}$
            \fi}

\def\CO#1{\ifnum#1=0                    
           \ifmmode \mbox{\rm CO}
           \else {\rm CO}
           \fi
          \else
           \ifnum#1<15
            \ifmmode ^{#1}\mbox{\rm CO}
            \else $^{#1}${\rm CO}
            \fi
           \else
            \ifmmode \mbox{\rm C}^{#1}\mbox{\rm O}
            \else {\rm C}$^{#1}${\rm O}
            \fi
           \fi
          \fi}

\def\COp{\ifmmode \mbox{\rm CO}^+           
         \else {\rm CO}$^+$                 
         \fi}

\def\CS#1{\ifnum#1=0                    
           \ifmmode \mbox{\rm CS}
           \else {\rm CS}
           \fi
          \else
           \ifnum#1<15
            \ifmmode ^{#1}\mbox{\rm CS}
            \else $^{#1}${\rm CS}
            \fi
           \else
            \ifmmode \mbox{\rm C}^{#1}\mbox{\rm S}
            \else {\rm C}$^{#1}${\rm S}
            \fi
           \fi
          \fi}

\def\HCOp{\ifmmode \mbox{\rm HCO}^+          
          \else {\rm HCO}$^+$                
          \fi}
\def\Hthreep{\ifmmode \mbox{\rm H}_3^+         
             \else {\rm H}$_3^+$               
             \fi}

\def\Htwo{\ifmmode \mbox{\rm H}_2              
          \else {\rm H}$_2$                    
          \fi}

\def\HtwoO{\ifmmode \mbox{\rm H}_2\mbox{\rm O} 
           \else {\rm H}$_2${\rm O}            
           \fi}

\def\ion#1#2{\ifmmode \mbox{{\rm #1}}\,\mbox{{\sc #2}} 
        \else {\rm #1}$\,${\sc #2}
        \fi}

\def\rec#1#2{\if#2a                            
              \ifmmode \mbox{{\rm #1}}\alpha   
              \else {\rm #1}$\alpha$
              \fi
             \fi
             \if#2b
              \ifmmode \mbox{{\rm #1}}\beta
              \else {\rm #1}$\beta$
              \fi
             \fi
             \if#2g
              \ifmmode \mbox{{\rm #1}}\gamma
              \else {\rm #1}$\gamma$
              \fi
             \fi}

\newcommand{\tabref}[1]{Table~\protect\ref{#1}}
\newcommand{\figref}[1]{Fig.~\protect\ref{#1}}

\newcommand{\eqref}[1]{Eq.~$\left(\protect\ref{#1}\right)$}

\begin{document}

\begin{frontmatter}



\title{Lessons from lensed Lyman break galaxies: can dusty 
Lyman break galaxies produce the submillimetre counts and background?}


\author{Paul P. van der Werf},
\ead{pvdwerf@strw.leidenuniv.nl}
\ead[url]{http://www.strw.leidenuniv.nl/$\tilde{\ }$pvdwerf}
\author{\hspace{-0.5cm}Kirsten Kraiberg Knudsen},
\author{Ivo Labb\'e}, \& 
\author{Marijn Franx}
\address{Leiden Observatory, P.O.~Box~9513, NL~2300~RA~Leiden, The Netherlands}

\begin{abstract}

Can the submillimetre counts and background be produced by applying a
locally derived extinction correction to the population of Lyman break
galaxies? We investigate the submillimetre emission of two strongly
lensed Lyman break galaxies ($\sou{MS}{1512{+}36}$-cB58 and
$\sou{MS}{1358{+}62}$-G1) and find that the procedure that is used to
predict the submillimetre emission of the Lyman break galaxy population
overpredicts the observed $850\mum$ fluxes by up to a factor of 14.
This result calls for caution in applying
local correlations to distant galaxies. It also shows that large
extinction corrections on Lyman break galaxies should be viewed with
skepticism. It is concluded that the Lyman break galaxies may
contribute to the submillimetre background at the 25 to 50\% level.
The brighter submillimetre galaxies making up the rest of the
background are either not detected in optical surveys, or if they are
detected, their submillimetre emission cannot be reliably estimated
from their rest-frame ultraviolet properties.

\end{abstract}


\end{frontmatter}


\section{Submillimetre emission of Lyman break galaxies}
\label{sec.LBGs}

Measurements of the cosmic star formation density (SFD) based on
surveys for Lyman break galaxies (LBGs) are affected by extinction,
and attempts to correct for this effect lead to a substantial upwards
revision of the SFD \citep{Meureretal99}.
Since the light absorbed in the ultraviolet (UV) is reradiated in the 
far-infrared (FIR), LBGs affected by extinction
must emit FIR radiation, which will contribute to the submillimetre
background. 

However, the lack of plausible counterpart LBGs in
deep submillimetre surveys with SCUBA
at $850\mum$ \citep{Hughesetal98} has given rise
to the view that the submillimetre galaxies form a separate
population which is not represented in LBG samples in the first place, 
a point of view which has been the subject of
considerable debate. We here discuss the relation between the LBGs and the
submillimetre galaxies in the light of existing data on submillimetre
source counts and background radiation, and recent measurements of
submillimetre emission from LBGs, including lensed LBGs. 

The expected submillimetre emission from the population of LBGs has
recently been estimated based on the observed (if not entirely
understood) correlation between the spectral index $\beta$ in the
UV (defined by the relation
$f_\lambda\propto\lambda^\beta$) and the ratio of FIR to UV flux in a
sample of local galaxies observed with the IUE
satellite \citep{Meureretal99}. Applying this relation to the LBG
population, \citet{AdelbergerSteidel00} 
found that the submillimetre counts and integrated 
background can be accounted for. As noted by these authors, these
estimates are still uncertain, since 
the validity of the $\beta$-FIR/UV correlation
at high redshift has not been established, and important
ingredients in the analysis such as the distribution of $\beta$ values and
its dependence on magnitude, and the luminosity function at faint
magnitudes are poorly constrained. 
It is therefore necessary to verify this analysis by
direct submillimetre observations of LBGs.

The predicted $850\mum$ fluxes for most LBGs based on their UV
properties \citep{AdelbergerSteidel00} are $1\un{mJy}$ or less,
which is too faint for current instrumentation. However, some of the
reddest LBGs are predicted to produce detectable $850\mum$ emission
and a sample of such LBGs can be selected based on their observed rest-frame UV
color and magnitude. However, a deep search for $850\mum$ emission from a
sample of 8 LBGs predicted to have $S_{850}>1\un{mJy}$ produced only
one detection; it was concluded that the $\beta$-FIR/UV correlation
overpredicted the FIR emission by at least a factor of
two \citep{Chapmanetal00}. Recently, it has been argued, based on an
analysis of structure in the faint $850\mum$ emission of the Hubble
Deep Field (HDF),
that the LBGs contribute at least 25\% but at most 50\% of the submillimetre 
background \citep{Peacocketal00}. These results indicate that the
LBGs do contribute to the submillimetre background in a significant
way, but that they do not produce a dominant contribution.

\begin{table}[t]
\caption{Predicted (from UV color and magnitude) and observed
submillimetre emission of lensed LBGs. Luminosities (not corrected for
gravitational amplification) have been derived
for $H_0=50\kms\pun{Mpc}{-1}$ and $q_0=0.5$. The other results do not
depend on cosmology. Upper limits represent $3\sigma$.
\label{tab.results}}
\vspace{0.4cm}
\begin{center}
\footnotesize
\begin{tabular}{|c|c|c|}
\hline
{} &\raisebox{0pt}[13pt][7pt]{$\sou{MS}{1512{+}36}$-cB58} &
\raisebox{0pt}[13pt][7pt]{$\sou{MS}{1358{+}62}$-G1}\\
\hline
& & \\[-20pt]
amplification factor & $\sim50$ & $\sim10$ \\[-8pt]
$\beta$ (observed) & $-0.74\pm0.1$ & $-1.63\pm0.1$ \\[-8pt]
$A_{1600}$ (predicted) & $3.0\pm0.2$ & $1.2\pm0.2$ \\[-8pt]
$\qu{S}{FIR}/\qu{S}{1600}$ (predicted) & 18 & 2.4 \\[-8pt]
$\qu{L}{FIR}$ (predicted) & $3.9\cdot10^{13}\Lsun$ &
$3.3\cdot10^{12}\Lsun$ \\[-8pt]
$S_{850}$ (predicted) & $58\un{mJy}$ & $5\un{mJy}$ \\[-8pt]
$S_{850}$ (observed)  & $4.2\pm0.9\un{mJy}$ & $<4\un{mJy}$ \\[-8pt]
$\qu{L}{FIR}$ (derived) & $2.8\cdot10^{12}\Lsun$ &
$<2.6\cdot10^{12}\Lsun$ \\[-8pt]
$\qu{S}{FIR}/\qu{S}{1600}$ (derived) & 1.3 & $<1.9$ \\[-8pt]
$A_{1600}$ (derived) & 0.8 & $<1.0$ \\[5pt]
\hline
\end{tabular}
\end{center}
\vspace{0.4cm}
\end{table}

\section{Lessons from strongly lensed Lyman break galaxies}
\label{sec.results}

\begin{figure}[t]
\begin{center}
\includegraphics*[width=9cm]{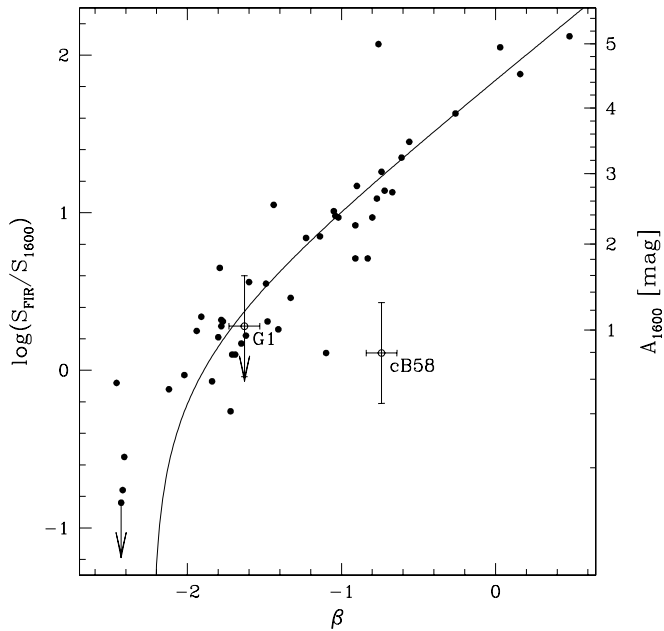}
\end{center}
\caption{The relation between $\qu{S}{FIR}/\qu{S}{1600}$ and $\beta$
for local UV-selected galaxies (filled circles) with the
best-fitting parametrization \protect\citep{Meureretal99}, and the positions of
cB58 and G1 with respect to this relation. \label{fig.beta}}
\end{figure}

Submillimetre observations of LBGs are significantly
easier if strongly lensed LBGs are targeted. We have used SCUBA on the
JCMT to observe two strongly lensed LBGs: the object cB58 at $z=2.72$ lensed by
the cluster $\sou{MS}{1512{+}36}$ \citep{Yeeetal96} and the
object G1 at $z=4.92$ lensed by the cluster 
$\sou{MS}{1358{+}62}$ \citep{Franxetal97}. Rest-frame UV colors
indicate significant reddening in both of these 
objects \citep{Ellingsonetal96,Soiferetal98}. In cB58, the Balmer
decrement also indicates the presence of extinction \citep{Teplitzetal00}
and in fact cB58 is
among the reddest quartile of LBGs \citep{Steideletal99}. 
For both galaxies, the method of predicting the FIR emission
based on the UV spectral slopes 
$\beta$ \citep{Bechtoldetal97,Pettinietal00,Soiferetal98} combined
with the flux densities at $1600\un{\AA}$ in the rest frame implies
strong $850\mum$ emission (\tabref{tab.results}).
The results of our SCUBA measurements are given in
\tabref{tab.results}. The object cB58 is detected at the $4.7\sigma$
level; this detection is confirmed by independent measurements at
$250\un{GHz}$ with the MAMBO instrument at the IRAM $30\un{m}$ 
telescope \citep{Baker00}. The object G1 was not detected; the
quoted upper limit comes from a combination of SCUBA photometry at the
position of the brightest knot and SCUBA mapping of the
lensing cluster. In both cases the procedure of predicting the
submillimetre flux from the observed color and magnitude in the
rest-frame UV \citep{AdelbergerSteidel00} overpredicts the
submillimetre emission. The
magnitude of the discrepancies is illustrated in \figref{fig.beta}.
While for G1 the discrepancy is not significant, given the scatter in
the $\beta$-FIR/UV relation, the discrepancy of a factor 14
for cB58 is highly significant.
These estimates should be independent of gravitational lensing if the
UV and FIR radiation originate from the same region, which is a
requirement for a meaningful prediction of FIR radiation fom UV properties.

Since for cB58 the discrepancy is so large, it cannot be attributed to
uncertainties, whether in the observations or in the analysis. For
instance, photometric errors (which would affect the determination of
$\beta$) are of minor importance, since this galaxy is lensed and
therefore relatively bright. The largest uncertainty in the predicted
$850\mum$ flux comes from converting the derived FIR luminosity into a
monochromatic flux density at the observing wavelength, which involves
the adoption of a FIR spectral energy distribution
(SED). For instance, exceptionally hot dust emission would produce a relatively
small $850\mum$ flux. However, the range of plausible SEDs
based on data of local galaxies 
introduces at most a factor 2 uncertainty in the predicted $850\mum$
flux density \citep{AdelbergerSteidel00}.

A final cause of uncertainty
in the analysis is introduced by differential lensing, if the
effective amplification factor of cB58 in the UV is a 
factor 14 higher than that
in the FIR\null. 
A large discrepancy can only be introduced 
if most of the UV emission comes from the most strongly amplified
portions of the source near the caustic, while most of the FIR
emission comes from more weakly lensed regions. This situation would
require a very different distribution of FIR and UV emission, which is
not impossible, as shown by the example of the nearby 
starburst merger $\NGC{4048{-}4039}$ where the dominant
region of obscured star formation as revealed by SCUBA mapping at
$850\mum$ \citep{VanDerWerfetal01} is spatially separated from the light
dominating the blue and ultraviolet fluxes. However, an
extinction correction based on UV properties of one region of a galaxy
will not be able 
to predict the submillimetre emission in a completely different
region of the system. Therefore, if differential lensing plays a major
role, the physical basis of using the $\beta$-FIR/UV correlation disappears.
For cB58, a lensing model based on new HST data shows
that the amplification factor is at least a factor of 5 at every
position, and that the intensity-weighted amplification factor in the UV is
about a factor of 25 (on both sides of the fold in the arc, so that the total
amplification is approximately a factor 50). 
Thus differential lensing can account for a
discrepancy of at most a factor of 5, but probably a lot less, since
UV and FIR emission should have a similar morphology for the
$\beta$-FIR/UV correlation to work.

\section{Discussion and conclusions}

These results demonstrate that attempts to produce the submillimetre counts and
background based on an extinction correction applied to the LBG 
population \citep{AdelbergerSteidel00} are fraught with considerable
uncertainty, and fail in the case of the two lensed LBGs discussed
here. While our sample is small, the
results argue against the validity of the low redshift $\beta$-FIR/UV
correlation in the case of LBGs. 
Even if the local correlation were valid for high redshift
galaxies as well, it still would not be able to produce the brighter
submillimetre galaxies. These objects have luminosities
putting them in the class of the ultraluminous infrared galaxies
(ULIGs), which do not follow the $\beta$-FIR/UV correlation, as shown
by recent HST-STIS data of a sample of nearby ULIGs. Since these
galaxies already account for $\sim50\%$ of the submillimetre
background, it is not possible for the LBGs to produce a dominant
fraction (let alone all) of the submillimetre background. A more
likely situation is that the LBGs account for 25 to 50\% of the
submillimetre background as indicated by the faint structure in the
HDF at $850\mum$ \citep{Peacocketal00}, but that the dominant part of
the submillimetre background is made by a small number of ULIGs, which
are either not detected in LBG surveys, or if they are detected,
cannot be reliably corrected for extinction, in the same way that
local ULIGs cannot be extinction corrected based on UV data.

In summary therefore, these results support the view that 
the brighter $850\mum$ galaxies making at least
50\% of the submillimetre background, form a population 
which cannot be reliably reproduced by extinction corrections applied
to the LBG population. 






\bibliographystyle{elsart-harv}

\bibliography{%
strings,%
cB58,%
HDFN,%
IRbackground,%
LBGs,%
LMCSMC,%
MS1358+62,%
NGC4038-4039,%
SFhistory,%
submmsurveys%
}

\end{document}